\documentclass[prl,aps,floats,showpacs,amsmath,amssymb,preprint]{revtex4}
\usepackage{bm,graphicx,epsf}
\usepackage{bm,graphicx}
\newcommand{\be}{\begin{equation}}
\newcommand{\ee}{\end{equation}}   
\newcommand{\bea}{\begin{eqnarray}}
\newcommand{\eea}{\end{eqnarray}}
\newcommand{\phrl}[1]{Phys.~Rev.~Lett. {\bf #1}}
\newcommand{\phrb}[1]{Phys.~Rev.~B {\bf #1}}

\newcommand{\jpsj}[1]{J.~Phys.~Soc.~Jpn.{\bf #1}}
\newcommand{\bib}{\bibitem}
\newcommand{\lb}{\left[}
\newcommand{\rb}{\right]}
\newcommand{\lp}{\left(}
\newcommand{\rp}{\right)}

\newcommand{\q}{{\bf q}}
\renewcommand{\k}{{\bf k}}
\newcommand{\Q}{{\bf Q}}

\newcommand{\F}{{\cal F}}

\begin{document}
\title{ Nature and symmetry of the order parameter of the\\ noncentrosymmetric superconductor $Li_2Pt_3B$}
\author{Soumya P. Mukherjee and Tetsuya Takimoto}
\affiliation{Asia Pacific Center for Theoretical Physics, Hogil Kim Memorial building 5th floor, POSTECH, Hyoja-dong, Namgu, Pohang 790-784, Korea } 

\date{\today}


\begin{abstract} 

 The nature and symmetry of the superconducting gap function in the noncentrosymmetric superconductor (NCS) $Li_2Pt_3B$, even many years after its discovery, 
appears to be full of contradictions. In this letter based on the existing band structure calculations we find that owing to the considerable nesting near the Fermi surface and 
the enhanced d-character of the relevant bands that cross the Fermi level,the system gets somewhat strongly correlated. Considering the effect of the onsite Coulomb repulsion 
on the pairing potential perturbatively, we extract possible superconducting transition. The strong normal spin fluctuation gives rise to a singlet dominant gap function with 
accompanying sign change. Thus our theory predicts a $s_\pm$ wave gap function with line nodes as the most promising candidate in the superconducting state. 
  
\end{abstract}

\maketitle
The occurrence of superconductivity in compounds without spatial inversion symmetry is one of the most active fields of research now a days. Inversion symmetry breaking leads to
many new  interesting effects in the superconducting state. The discovery of superconductivity in $Li_2Pd_3B$ \cite{Togano} and subsequently the experiments in the pseudo-binary 
complete solid solution $Li_2(Pd_{1-x}Pt_x)_3B$, $x=0\sim1$ \cite{Badica} attracted much attention. Many experimental as well as theoretical works reported since 
then. The end compounds $Li_2Pd_3B$ $(x=0)$ and $Li_2Pt_3B$ $(x=1)$ \cite{Nishiyama1,Nishiyama2,Takeya1,Takeya2,Yuan,Hafliger} were also studied intensively and compared. It is now 
established that the superconductivity in $Li_2Pd_3B$ is phonon mediated s-wave type. The presence of Hebel-Slichter peak in the Nuclear spin-lattice relaxation rate measurement
\cite{Nishiyama1,Nishiyama2}, low temperature behavior of the specific heat \cite{Takeya1,Takeya2}, penetration depth \cite{Yuan} etc. as well as NMR Knight shift data strongly 
support this conclusion. On the other hand, the nature and symmetry of the gap function of the compound $Li_2Pt_3B$ is still debatable. Similar experiments performed on this compound 
\cite{Nishiyama2,Takeya2,Yuan} suggest the presence of line nodes in the superconducting state. The NMR Knight shift, often used to distinguish the spin state of superconductivity 
between singlet and triplet, is almost temperature independent even below $T_c$.This behavior is also interesting and deserve special attention. 

In this letter, based on the existing band structure calculations \cite{Chandra,Lee} we find that there exists considerable nesting between the Fermi surfaces and enhanced d-character 
of the relevant bands that cross the Fermi level. These two effects lead to a stronger electron correlation in $Li_2Pt_3B$ than in $Li_2Pd_3B$ \cite{Yokoya}. By treating this 
correlation perturbatively we estimate the effect of spin fluctuations on the stability of superconducting state \cite{Takimoto1}. Owing to the large usual spin fluctuation the singlet 
gap function becomes much stronger than the triplet gap function. The singlet gap function belongs to $A_1$ representation with sign change between two branches of the Fermi surface. 
Thus our theory suggests a singlet dominant $s_\pm$ wave gap function with line nodes as the most promising candidate for the superconducting state of $Li_2Pt_3B$. This prediction 
also explains most of the experiments. We also calculate the behavior of uniform spin susceptibility below $T_c$ and comment on the apparent mismatch with the experiment. 

The crystal structure of the compound $Li_2Pt_3B$ is simple cubic (with point group O) and isostructural with the compound $Li_2Pd_3B$. The only difference between them is in the 
mass of the central Pt and Pd atoms. However this gives rise to some significant observable effects \cite{Chandra,Lee}. For Pt compound there is an enhancement of d-character of the 
bands that cross the Fermi level. This enhancement of the d-character is reflected in the increased DOS at the Fermi level. Considering these, we can construct a minimal model 
Hamiltonian $(H=H_0 + H_1)$ of $Li_2Pt_3B$ which is given by the Hubbard model with an antisymmetric spin-orbit (SO) coupling term, where
\be
H_{0}=\sum_{\k \sigma \sigma'} \lp \lb \varepsilon_{\k}-\mu\rb \hat{\sigma}_{o}+ {\bf g}_{\k}.\hat{\sigma}\rp_{\sigma \sigma'} c^\dagger_{\k \sigma} c_{\k \sigma'},
\label{H0}
\ee
and $H_{1}=U\sum_{i} n_{i \uparrow}n_{i \downarrow}$.
Here $c_{\k \sigma}$ and $c^\dagger_{\k \sigma}$ denotes the annihilation and creation operators of an electron with momentum $\k$ and spin $\sigma$. $\varepsilon_{\k}$ is the 
dispersion of electrons and $\mu$ the chemical potential. ${\bf g}_{\k}=-{\bf g}_{-\k}$ denotes the effective anti-symmetric SO coupling which breaks the inversion symmetry. In $H_1$, 
U is the screened on-site interaction. The dispersion of electrons $\varepsilon_{\k}$ is constructed by the tight-binding method including upto fourth-neighbor hopping in the 
three-dimensional simple cubic lattice. 
\bea
\nonumber
&& \varepsilon_{\k}=2t_{1}(\cos(k_x)+\cos(k_y)+\cos(k_z)) +4t_2(\cos(k_x) \cos(k_y)+ \cos(k_y)\cos(k_z)\\ 
&& +\cos(k_z)\cos(k_x))+ 8t_3  \cos(k_x) \cos(k_y)\cos(k_z)+2t_4 (\cos(2k_x)+\cos(2k_y)+\cos(2k_z))
\label{epsilon}
\eea
\begin{figure}
\includegraphics[height=8cm,width=6cm,angle=270]{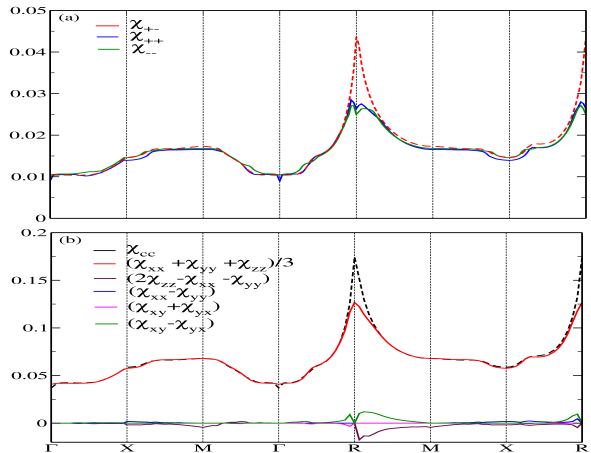}
\caption{(Color online) Signature of nesting.(a) Momentum dependence of interband and intraband susceptibilities showing nesting at ${\bf Q}=(\pi,\pi,\pi)$.(b)The comparison of the 
relative magnitudes of normal and anomalous spin fluctuations. The charge fluctuation $\chi_{cc}$ and normal spin fluctuation $\frac{1}{3}(\chi_{xx}+\chi_{yy}+\chi_{zz})$ contribute 
strongly compared to the anomalous spin fluctuations.The spin susceptibility also get enhanced at the R point (details in text).}
\label{fig:nesting}
\end{figure}
The SO coupling term appropriate for the point group is given as ${\bf g}_{\k}=g(\sin(k_x),\sin(k_y),\sin(k_z))$. The values of the parameters $\lp t_1,t_2,t_3,t_4,g,\mu \rp$ 
are chosen to be $\lp 1.0,-0.03,-0.88,-0.03,0.5,0.02\rp$ as Fermi surface obtained by the band structure calculation are reproduced. One can diagonalize $H_{0}$ to get the eigen-energies 
$\varepsilon_{\k\pm}=\varepsilon_{\k}\pm |{\bf g}_{\k}|-\mu$. The spin degeneracy is now removed and $\varepsilon_{\k \pm}$ gives us the energy of the two helicity bands. The Fermi 
surfaces corresponding to the helically splitted bands consist of three major branches. One electron pocket around the $\Gamma$ point, one hole pocket around the R point and other 
remaining parts forming a cage like structure with neck and mouth along $\Gamma-X $ direction. The corresponding parts of the Fermi surfaces of the different helicity bands are shifted 
from each other depending on the magnitude of $g$. Owing to the smallness of the parameters $t_2,t_4$ and $\mu$ there appears a large nesting with the nesting vector 
${\bf Q}=(\pi,\pi,\pi)$ connecting between the cage like larger portion of the Fermi surface of the negative helicity band $\varepsilon_{\k -}$ and the similar cage like Fermi surface of
the positive helicity band $\varepsilon_{\k +}$. Therefore this nesting gives rise to a sharp pick at the R point in the momentum dependence of 
$\chi_{+-}(\q)= \frac{1}{8N_0}\sum_{\k} \frac{f(\varepsilon_{\k -})-f(\varepsilon_{\k+\q +})}{\varepsilon_{\k+\q +}-\varepsilon_{\k -}}$ where nesting condition is satisfied. There also 
exists partial nesting between Fermi surfaces around $\Gamma-$ and R-points of both helicity bands but for positive helicity band it's larger as shown in Fig.~\ref{fig:nesting}(a).
\begin{figure}
\includegraphics[height=7cm,width=4cm,angle=270]{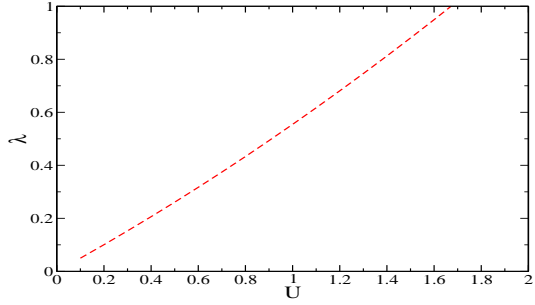}
\caption{(Color online) The variation of maximum eigenvalue with U.}
\label{fig:eigenval}
\end{figure}
The most general form of the matrix gap function is $\hat{\Delta}_\k=\lb\Psi(\k)\hat{\bf{\sigma}}_0+{\bf d(\k)}.\hat{\sigma}\rb i\hat{\sigma}_y$. Here $\Psi(\k)$ is the 
singlet gap function and $\bf d(\k)$ is the triplet $\bf d$-vector. In NCS superconductor triplet component with $|{\bf d(\k)}.{\bf g}_{\k}|=|{\bf d(\k)}||{\bf g}_{\k}|$
survive the pinning from the antisymmetric SO coupling \cite{Sigrist}. So one can write $\bf d(\k)=\phi(k){\bf g}_{\k}$ where $\phi(\k)$ having the same symmetry of momentum dependence 
as $\Psi(\k)$.
With all these we can define the normal $\hat{G}(\k,i \omega_n)$ and anomalous $\hat{F}(\k,i \omega_n)$ matrix Green's functions as below,
\be
\hat{G}(\k,i \omega_n)=G_{+}(\k,i \omega_n) \hat{\sigma}_0+G_{-}(\k,i \omega_n)\tilde{\bf g}_{\k}.\hat{\bf{\sigma}},
~\hat{F}(\k,i \omega_n)=\lb F_{+}(\k,i \omega_n) \hat{\sigma}_0+F_{-}(\k,i \omega_n)\tilde{\bf g}_{\k}.\hat{\bf{\sigma}}\rb i \hat{\sigma}_y
\label{Green}
\ee  
here $\tilde{\bf g}_{\k}={\bf g_\k}/|{\bf g_\k}|$. $G_{\pm}$ and $F_{\pm}$ are given as, 
\bea
\nonumber
G_{\pm}(\k,i \omega_n)= \frac{1}{2}\lp \frac{-i \omega_n-\varepsilon_{\k +}}{{\omega_n}^2+E_{\k +}^2}\pm \frac{-i \omega_n-\varepsilon_{\k -}}{{\omega_n}^2+E_{\k -}^2}\rp, 
F_{\pm}(\k,i \omega_n)= \frac{1}{2}\lp \frac{\Delta_{\k +}}{{\omega_n}^2+E_{\k +}^2}\pm \frac{\Delta_{\k -}}{{\omega_n}^2+E_{\k -}^2}\rp ,
\eea
here $\Delta_{\k \pm}=\Psi(\k)\pm \phi(\k)|{\bf g_{\k}}|$ and $E_{\k \pm}= \sqrt{\varepsilon_{\k \pm}^2+\Delta_{\k \pm}^2}$. Within the weak coupling theory of superconductivity only 
static susceptibility is required. We start by defining the dynamical susceptibility as,
\be
\chi_{\alpha \beta}(\q,i\Omega_n)= \int^{1/T}_0{d\tau e^{i \Omega_{n}\tau}\langle T_\tau [S^\alpha_\q(\tau) S^\beta_{-\q}(0)]\rangle}
\ee
here $\langle ...\rangle$ denotes thermal average, $T_\tau$ imaginary time ordering and $\Omega_n$ are the Bosonic Matsubara frequencies. The charge (spin) operators with wave vector 
$\q$ is defined as,
\be
S^c_{\q}=\frac{1}{2}\sum_{\k \sigma} c^\dagger_{\k \sigma} c_{\k+\q \sigma},
~~ S^\alpha_{\q}=\frac{1}{2}\sum_{\k \sigma \sigma'} \sigma^\alpha_{\sigma \sigma'}c^\dagger_{\k \sigma} c_{\k+\q \sigma'}
\ee 
With all these, the matrix elements of the static spin susceptibilities $\chi_{\alpha \beta}(\q)$ for $\alpha,\beta=c,x,y,z$ is found to be,
\be
\chi_{\alpha \beta}(\q)= \frac{1}{8N_0}\sum_{\k}\sum_{\xi \zeta} \Gamma^{\alpha \beta}_{\xi \zeta} (\k;\q) \frac{f(\varepsilon_{\k \xi})-f(\varepsilon_{\k+\q \zeta})}
{\varepsilon_{\k+\q \zeta}-\varepsilon_{\k \xi}},
\ee
where $f(\varepsilon)$ is the Fermi distribution function and the function $\Gamma^{\alpha \beta}_{\xi \zeta}$ is obtained as,
\be
\Gamma^{\alpha \beta}_{\xi \zeta}(\k;\q)= \delta_{\alpha,\beta} (1-\xi \zeta \tilde{\bf g}_{\k}.\tilde{\bf g}_{\k+\q})
+\xi \zeta(\tilde{g}_{\k \alpha}\tilde{g}_{\k+\q \beta}+ \tilde{g}_{\k \beta}\tilde{g}_{\k+\q \alpha}) 
-\epsilon_{\alpha \beta \gamma} i(\xi \tilde{g}_{\k+\q \gamma}-\zeta \tilde{g}_{\k \gamma}).
\ee
Similarly the charge fluctuation in the normal state i.e. $ \chi_{cc}(\q)$ is obtained with the replacement of $\Gamma^{\alpha \beta}_{\xi \zeta}(\k;\q) $ by 
$\Gamma^{cc}_{\xi \zeta}(\k;\q)$ where $\Gamma^{cc}_{\xi \zeta}(\k;\q)= 1+\xi \zeta \tilde{\bf g}_{\k}.\tilde{\bf g}_{\k+\q}$. The susceptibilities between spin and charge operators
$\chi_{c \alpha}(\q)$ and $\chi_{\alpha c}(\q)$ all vanishes for the static case. We calculate all the susceptibility components and examine the property of spin fluctuations. The usual 
spin fluctuation $\frac{1}{3}(\chi_{xx}+\chi_{yy}+\chi_{zz})$ with the momentum dependence of $q^2-$ type is also present in the centrosymmetric cubic system. Other symmetric spin 
fluctuations $(2\chi_{zz}-\chi_{xx}-\chi_{yy}), (\chi_{xx}-\chi_{yy})$ and $(\chi_{\alpha \beta}+\chi_{\beta \alpha})$ with $\alpha \neq \beta$ having momentum dependence 
$2q^2_z-q^2_x-q^2_y, q^2_x-q^2_y$ and $q_{\alpha}q_{\beta}(\alpha \neq \beta)-$ types respectively are special to the cubic noncentrosymmetric case \cite{Takimoto1}. Along with these the 
anti-symmetric spin fluctuations $i(\chi_{\alpha \beta}-\chi_{\beta \alpha})$ with $\alpha \neq \beta$ with momentum dependence $q_{\gamma}(\gamma \neq \alpha \neq \beta)-$ type are 
also present. In Fig.~\ref{fig:nesting}(b) we compare relative strengths of the charge and usual spin fluctuation together with the anomalous spin fluctuations along symmetrical lines. 
Later we will see that the largeness of the usual spin fluctuation is responsible for the largeness of the singlet gap function as triplet gap function whose magnitude is much smaller 
than the singlet one are induced by the antisymmetric spin fluctuations. 
\begin{widetext}
\begin{figure}
\vspace{-0.6cm}
\includegraphics[height=4cm,width=16cm,angle=180]{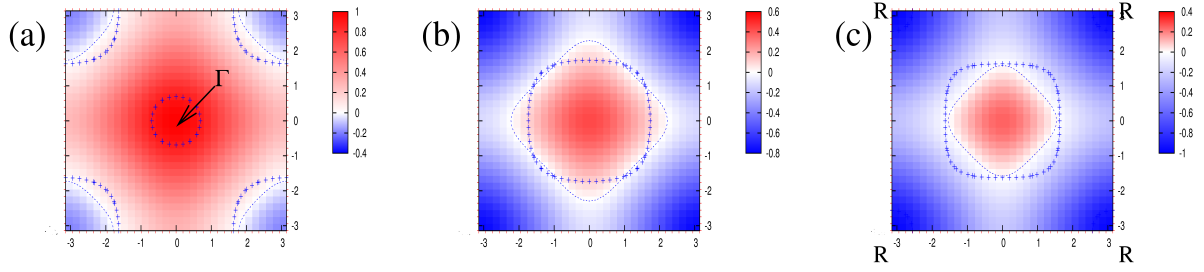}
\caption{(Color online) The singlet gap function at three $k_z$ values; (a) $k_z=0.0$, (b)$k_z=0.7 \pi$, and (c)$k_z= \pi $. Gap function varies from positive-maximum (Red) and to the 
negative-maximum (blue) following the scale attached with each figure. The blue line with $+$ sign shows the Fermi surface of the positive helicity band while the dotted blue line 
denotes the nodes of the gap function (details in text).}
\label{fig:psi}
\end{figure}
\end{widetext}

Treating the interaction term perturbatively and following the standard procedure \cite{Takimoto2,Tada,Yanase} we arrive at the following superconducting gap equation,
\be
\left( \begin{array}{cccc}
\Psi(\k)\\ d_x(\k)\\  d_y(\k)\\  d_z(\k) \end{array} \right) 
=\frac{1}{N_0}\sum_{\q} \label{gap_equ}
\left( \begin{array}{rrrr}
V_{ss}(\q) & V_{sx}(\q) & V_{sy}(\q) & V_{sz}(\q) \\
V_{xs}(\q) & V_{xx}(\q) & V_{xy}(\q) & V_{xz}(\q) \\
V_{ys}(\q) & V_{yx}(\q) & V_{yy}(\q) & V_{yz}(\q) \\
V_{zs}(\q) & V_{zx}(\q) & V_{zy}(\q) & V_{zz}(\q) \end{array} \right)
\left( \begin{array}{cccc}
\F_{s}(\k-\q) \\ \F_{x}(\k-\q) \\ \F_{y}(\k-\q) \\ \F_{z}(\k-\q) \end{array} \right),
\ee
here $V_{\zeta \eta}$ with $(\zeta, \eta)=(s,x,y,z)$ denotes the pairing potential arising from the corresponding fluctuation exchange and they are expressed as below,
\bea
\nonumber
&&V_{ss}(\q)=-U-U^2\lb\chi_{xx}(\q)+\chi_{yy}(\q)+\chi_{zz}(\q)-\chi_{cc}(\q)\rb\\ \nonumber
&&V_{\zeta \zeta}(\q)=U^2 \lb \chi_{cc}(\q)+\chi_{\eta \eta}(\q)+\chi_{\delta \delta}(\q)-\chi_{\zeta \zeta}(\q)\rb \\ \nonumber
&&V_{\zeta \eta}(\q)= V_{\eta \zeta}(\q)=-U^2 \lb \chi_{\zeta \eta}(\q)+\chi_{\eta \zeta}(\q)\rb \\
&&V_{s \zeta}(\q)= -V_{\zeta s}(\q)=i U^2 \lb \chi_{\eta \delta}(\q)-\chi_{\delta \eta} (\q)\rb,
\label{pairing}
\eea
where $\zeta \neq \eta \neq \delta$. Here $\F_s$ and $\F_\alpha$ are the contributions from the anomalous Green's functions after frequency summation \cite{Takimoto2}. As we mentioned 
above the singlet component of the gap function arising from usual spin fluctuation dominates over triplet gap function which is induced by the small antisymmetric spin fluctuations. 
Eq.~(\ref{gap_equ}) reduces to the eigenvalue problem if we work at the transition temperature $T_c$. We fix $T_c$ arbitrarily at $0.02t_1$ and solve Eq.~(\ref{gap_equ}) for maximum 
eigenvalue. Fig.~\ref{fig:eigenval} gives us the critical value of the onsite interaction $U$ for superconductivity i.e. $U_c=1.675t_1$ , when maximum eigenvalue becomes unity. We thus 
get the momentum dependence of both singlet and triplet gap functions as the eigenfunctions of the maximum eigenvalue. From the momentum dependence of the gap function we conclude that 
the superconductivity belongs to the $A_1$ representation of the point group O.

In  Fig.~\ref{fig:psi} we present the contour plot of the singlet gap function at three $k_z$ values in the 1st Brillouin zone. The singlet gap function changes sign from positive (red)
to negative (blue) gradually moving from $\Gamma-$ to R-point and vanishes completely somewhere in between forming the nodal surface. Here we would like to mention that the gap function 
is strongest at either $\Gamma-$ or R-points although the nesting is rather weak here. On the other hand the cage like portion of the Fermi surface where we have most strongest nesting 
gives rise to weak gap function. It can be understood from a careful observation of Eq.~(\ref{pairing}). The summation of prefactor $\Gamma^{\alpha \alpha}_{+-}({\k,\Q})$ of susceptibility
$\chi_{\alpha \alpha}(\Q)$ vanishes for the spin-singlet pairing potential $V_{ss}(\Q)$ at the nesting vector ${\bf Q}=(\pi,\pi,\pi)$. Because of this, even strong interband nesting 
does not play any role in opening up the gap function. However, the pairing potential forms the gap on the Fermi surfaces around $\Gamma-$ and R-points, connected by sub-dominant 
nesting of $\chi_{++}(\Q)$. Thus the singlet gap function with opposite signs between these points opens up and line nodes can exist in between. In this figure we show the corresponding 
zeros of the positive helicity Fermi surface by the $+$ sign line and the dotted line denotes the exact location where the gap function vanish. In Fig.~\ref{fig:psi}(a) around $\Gamma-$ 
point the gap is positive maximum. In Fig.~\ref{fig:psi}(b) the strength of the negative gap function increases and finally, in Fig.~\ref{fig:psi}(c) we encounter the maximum negative 
value of the gap function at the corner R-point. Thus the gap function appears to be singlet $s_{\pm}$ type with accidental line nodes which is not allowed by symmetry, rather depend on 
the three-dimensional geometry of the Fermi surface.

We also calculate the temperature dependence of the susceptibility in the superconducting state. Within the weak coupling approximation neglecting the feedback effect we assume that 
the order parameter below $T_c$ follows the BCS temperature dependence $\hat{\Delta}(\k,T) = \hat{\Delta}(\k,0)\tanh(1.74\sqrt{\frac{T_c}{T}-1})$ at every momentum point. Using this 
gap function we calculate the uniform susceptibility below $T_c$ as follows \cite{Frigeri}. 
\be
\chi_{s}(T)=\sum_{\k} \lb \chi_0(\k,T) + \chi_{+}(\k,T) + \chi_{-}(\k,T) \rb,
\ee
where  $\chi_0(\k,T)$ , $\chi_{+}(\k,T)$ and $\chi_{-}(\k,T)$ are given as,
\be
\chi_0(\k,T)= \frac{1}{3}\sum_{\xi=\pm} [\lp 1-\xi \frac{\Delta_{\k +}\Delta_{\k -}+\epsilon_{\k +}\epsilon_{\k-}}{E_{\k +}E_{\k -}}\rp \times
\lp \frac{\tanh(\frac{E_{\k +}}{2T})+\xi \tanh(\frac{E_{\k -}}{2T})}{E_{\k +}+\xi E_{\k -}}\rp],
\ee
\be
\chi_{\pm}(\k,T)= 1/(12T\cosh^2(E_{\k \pm}/2T)),
\ee
\begin{figure}
\includegraphics[height=7cm,width=4cm,angle=270]{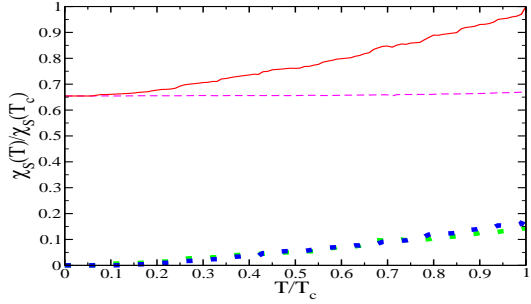}
\caption{(Color online) The variation of normalized spin susceptibility (solid red line) with temperature in the superconducting state. Also shown in the same figure the contributions 
from the Van-Vleck term (dashed magenta line) and the Pauli term of positive helicity band (green square)and negative helicity band (blue square), all scaled by $\chi_{s}(T_c)$.}
\label{fig:sus_temp}
\end{figure}
We plot in Fig.~\ref{fig:sus_temp} the contribution from the temperature independent Val-Vleck term $\chi_0(T)$, temperature dependent Pauli terms $\chi_+(T),\chi_-(T)$ and the
susceptibility $\chi_{s}$ (T) all normalized by $\chi_{s}(T_c)$ (red line). Although $\chi_{s}(T)/\chi_{s}(T_c)$ shows excellent agreement with earlier works \cite{Samokhin} but 
apparently contradicts the NMR Knight shift data \cite{Nishiyama2}. To explain the contradiction with the experiment, one can formulate a multi-orbital theory which captures the 
complicated band structure in more detail. Then the large contributions of the Van-Vleck term between $t_{2g}-$ and $e_g-$ orbitals for cubic system is expected and this will further 
reduce the deviation of the normalized susceptibility from normal state below $T_c$. This involves somewhat elaborate calculations and we leave this as a future problem.

In conclusion, we suggest that, in the noncentrosymmetric superconductor $Li_2Pt_3B$ considerable d-character of the bands near the Fermi energy and nesting of the Fermi surfaces 
give rise to weak correlation effect which can be treated perturbatively and this give rise to a singlet dominated (with negligible triplet component) $s_{\pm}$ kind of gap function 
with accidental line nodes arising from the Fermi surface geometry. The three-dimensional geometry of the Fermi surface and the nesting of the Fermi surface also play a crucial role in 
determining the nature of the gap function. We propose that angle-resolved photo emission spectroscopy and de Hass-van Alphen effect experiments may shed light on this nesting property 
of the Fermi surface and can be useful to study the properties of the superconducting state as well. We also calculate the susceptibility below $T_c$ and emphasize the importance of
orbital degeneracy of d-electron to explain the experimental data.  

The authors are grateful to Y. Yanase for fruitful stimulating discussions during his visit in APCTP.


\end{document}